\documentclass[conference]{IEEEtran}
\usepackage{graphicx}                       
\usepackage{amsmath,amsfonts,amssymb}       
\usepackage[hyphens]{url}
\usepackage[breaklinks=true]{hyperref}                
\usepackage{float}                          
\usepackage{lipsum}                         
\usepackage{newtxtext}                      
\usepackage{newtxmath}                      
\usepackage{multirow}                       
\usepackage{xcolor}                         
\usepackage{textcomp}                       
\usepackage{booktabs}                       
\usepackage{algorithm}                      
\usepackage{algpseudocode}                  
\usepackage{listings}                       

\usepackage{amsmath,amssymb,amsfonts}
\usepackage{tikz}
\usetikzlibrary{arrows,shapes,positioning,shadows,trees}
\tikzset{
	basic/.style  = {draw, text width=2cm, drop shadow, font=\sffamily, rectangle},
	root/.style   = {basic, rounded corners=2pt, thin, align=center,
		fill=blue!60},
	level 2/.style = {basic, rounded corners=2pt, thin,align=center, fill=blue!40,
		text width=3.1em},
	level 3/.style = {basic, thin, align=left, fill=blue!20, text width=3.5em}
}
\def\BibTeX{{\rm B\kern-.05em{\sc i\kern-.025em b}\kern-.08em
    T\kern-.1667em\lower.7ex\hbox{E}\kern-.125emX}}

\title{Automated Machine Learning: A Case Study on Non-Intrusive Appliance Load Monitoring}

\author{\IEEEauthorblockN{Armin Moin}
	\IEEEauthorblockA{\textit{Department of Computer Science} \\
		\textit{University of Colorado Colorado Springs}\\
		United States \\
		amoin@uccs.edu \textsuperscript{*}Corresponding author}
	~\\
	\and
	\IEEEauthorblockN{Ukrit Wattanavaekin}
	\IEEEauthorblockA{\textit{School of Computation, Information, and Technology (CIT)} \\
		\textit{Technical University of Munich}\\
		Munich, Germany \\
		ukritwk@gmail.com}
	~\\
	\and
	\IEEEauthorblockN{Alexandra Lungu}
	\IEEEauthorblockA{\textit{School of CIT} \\
		\textit{Technical University of Munich}\\
		Munich, Germany \\
		alexandra.lungu88@gmail.com}
	~\\
	\and
	\IEEEauthorblockN{Stephan R\"ossler}
	\IEEEauthorblockA{\textit{Software AG} \\
		Munich, Germany \\
		stephan.roessler@softwareag.com}
	~\\
	\and
	\IEEEauthorblockN{Stephan G\"unnemann}
	\IEEEauthorblockA{\textit{School of CIT and MDSI} \\
		\textit{Technical University of Munich}\\
		Munich, Germany \\
		s.guennemann@tum.de}
}

\begin{document}

\maketitle

\begin{abstract}
We propose a novel approach to enable Automated Machine Learning (AutoML) for Non-Intrusive Appliance Load Monitoring (NIALM), also known as Energy Disaggregation, through Bayesian Optimization. NIALM offers a cost-effective alternative to smart meters for measuring the energy consumption of electric devices and appliances. NIALM methods analyze the entire power consumption signal of a household and predict the type of appliances as well as their individual power consumption (i.e., their contributions to the aggregated signal). We enable NIALM domain experts and practitioners who typically have no deep data analytics or Machine Learning (ML) skills to benefit from state-of-the-art ML approaches to NIALM. Further, we conduct a survey and benchmarking of the state of the art and show that in many cases, simple and basic ML models and algorithms, such as Decision Trees, outperform the state of the art. Finally, we present our open-source tool, AutoML4NIALM, which will facilitate the exploitation of existing methods for NIALM in the industry.
\end{abstract}


\section{Introduction}
\label{sec:introduction}
Recently, Machine Learning (ML) has been applied to the application domain of \textit{Non-Intrusive Appliance Load Monitoring (NIALM)}, also known as \textit{Energy Disaggregation}. In fact, NIALM, which is an \textit{inverse problem}, is very similar to the Single-channel Source Separation problem in Physics and Signal Processing. However, audio signals tend to have a much higher temporal resolution than energy measurements, making our problem more challenging than that one \cite{Kolter+2010}. Here is the problem of NIALM in simple words: Let us assume that we have access to the information about the total electrical energy consumption of a household at certain points in time, e.g., every 15 minutes. NIALM is about finding out what electrical devices and appliances are likely to be in the house and estimating what their individual electrical energy consumption values have been over a certain period of time.

It is clear that one could achieve this goal through an \textit{intrusive} method by attaching smart meters to appliances in a house and measuring their electrical energy consumption levels. However, that intrusive solution would definitely cost more, both for the procurement of sensors and also their maintenance. Moreover, that solution would not be convenient for the end users (occupants) and/or owners of buildings since they would need to allow the installation of sensors (smart meters) in their houses. This may also raise privacy concerns \footnote{ It is worth mentioning that privacy concerns may still be the case in virtual sensing via NIALM. However, the concerns and inconveniences are not as bold as in the intrusive scenario.}. Therefore, the non-intrusive method through virtual sensing, also known as soft sensing, has become more popular. Generally, virtual sensing is usually performed either analytically by using some physical laws or empirically following a data-driven approach, such as using ML. In our context, only the latter is possible as we do not have any physical law that gives us the required information given the aggregated power signal (i.e., the total energy signal of a household). NIALM has a lot of applications, e.g., power grids (smart grids), which can better plan and balance their energy networks. Also, research (e.g., see  \cite{Amenta+2013}) showed that providing information about the power consumption of individual appliances to energy consumers often leads to more user awareness and usually results in higher energy savings.

This paper makes three contributions: (i) It proposes a novel approach based on Bayesian Optimization, with tool support to enable Automated Machine Learning (AutoML) for domain experts and practitioners in the field of NIALM. (ii) It conducts a survey and benchmarking of the state of the art and illustrates that a simple and basic ML approach, namely, Decision Trees, can outperform the more complex approaches to NIALM that exist in the literature. Our benchmarking is fair since we use the same dataset and settings for all experiments. So far, such a study is missing in the NIALM literature. (iii) It proposes new practical methods for NIALM, which employ Decision Trees - as one of the most basic ML algorithms - as well as Fully-Connected Neural Networks (FCNNs), also known as Multi-Layer Perceptrons (MLPs) - as the most basic architecture for Neural Networks. Further, it proposes another new method based on Gated Recurrent Units (GRU), a variation of Recurrent Neural Networks (RNN) that outperforms all the state-of-the-art approaches, including LSTMs, though - surprisingly - it does not yet perform on average better than Decision Trees.

The rest of this paper is structured as follows: In Section \ref{literature-survey-benchmarking}, we survey the literature of NIALM and briefly present our benchmarking results for existing approaches. Then, we propose our novel approach for enabling Automated Machine Learning in the field of NIALM and present our open-source tool for supporting that in Section \ref{proposed-approach}. Further, we illustrate our experimental results in Section \ref{experimental-results}, which also leads to an extended benchmarking that includes the new methods that we propose for NIALM applications. Finally, we conclude and suggest future work in Section \ref{conclusion-future-work}.

\section{Literature Survey \& Benchmarking}\label{literature-survey-benchmarking}
In this section, we survey the literature on NIALM. Existing approaches can be categorized from various points of view. One way is to group them into those approaches that require high-frequency data (e.g., see \cite{Sultanem1991,Hart1992,Shaw+1998,Laughman+2003,Lee+2004,Patel+2007,Berges+2009,LinTsai2014,Kramer+2014,Agyeman+2015,Gillis+2016}) and those that can work well even with low-frequency data (e.g., see \cite{Kolter+2010,Bergman+2011,Parson+2012,Shao+2013,Stankovic+2014,KellyKnottenbelt2015a,ElhamifarSystry2015,Zhong+2015,Elhamifar+2016,Zhao+2016,MauchYang2016,He+2018}). High-frequency here means sampling rates in the order of more than one measurement per second, i.e., above 1 Hz, whereas low-frequency means sampling rates in the order of one or less than one measurement per second, i.e., below 1 Hz. Note that the former (i.e., high-frequency data) would typically require expensive equipment that makes such approaches useless in practice as conventional counters of households cannot afford such rates. On the contrary, practical approaches that can work with conventional frequencies, e.g., even as low as one measurement per $15$ minutes or per hour (e.g., see \cite{Kolter+2010}), are interesting to us.

Another way of categorizing existing NIALM approaches is to group them into state-based and event-based approaches, though the border is sometimes not sufficiently clear. State-based approaches model each appliance or various states of each appliance as states of a Finite State Machine (FSM) or as signatures used in a dictionary. These approaches are usually computationally more expensive but, in the end, often more accurate. Two of the many examples in the literature are those proposed in \cite{MauchYang2016} (combined DNN-HMM model) and \cite{ElhamifarSystry2015} (Dictionary-based). However, the more classical approach is to focus on the changes in the aggregated signal using signal edges without considering various models for appliances or their states.

One example of this classic category is the so-called training-less approach based on Spectral Clustering proposed in \cite{Zhao+2016}. The authors, who were apparently not from the Machine Learning (ML) community, referred to their approach as non-ML. Nevertheless, their proposed approach is indeed an ML approach and is considered unsupervised ML, also known as clustering. However, we do not consider this approach - in its current stage - as practical either. First, like other event-based approaches, it has the drawback of being generally unable to properly disaggregate signals of appliances that are working simultaneously and appliances with varying loads (e.g., dishwashers or microwave ovens). In addition, one cannot use that unsupervised approach in practice without having a complementary mechanism to finally label the disaggregated signals with the name of the corresponding appliances.

Table \ref{tab:sota} illustrates our benchmarking results for existing approaches, which we find practical for NIALM applications based on the above-mentioned arguments. Our comprehensive benchmarking will come in Section \ref{experimental-results}.

\begin{table}[ht]
	\centering
	\caption{Benchmarking of existing state-of-the-art approaches to NIALM}
	\label{tab:sota}
	\begin{tabular}{|p{1cm}|p{3cm}|p{4cm}|}
		\hline
		\textbf{Rank} & \textbf{Paper} & \textbf{Algorithm} \\
		\hline
		1st & Kelly \& Knottenbelt 2015 \cite{KellyKnottenbelt2015a} & Long Short-Term Memory (LSTM) \\
		\hline
		2nd & Kelly \& Knottenbelt 2015 \cite{KellyKnottenbelt2015a} & Denoising Autoencoders (DAE) \\
		\hline
		3rd & Elhamifar \& Systry 2015 \cite{ElhamifarSystry2015} & Dictionary-Based Approach \\
		\hline
		4th & Mauch \& Yang 2016 \cite{MauchYang2016} & Combined Deep Neural Networks (DNN) and Hidden Markov Models (HMM) \\
		\hline
		5th & E.g., Reyes-Gomez et al. 2003 \cite{Reyes-Gomez+2003} & Factorial Hidden Markov Models (FHMM) \\
		\hline
		6th & Hart 1992 \cite{Hart1992} & Combinatorial Optimization (CO) \\
		\hline
	\end{tabular}
\end{table}

\section{Proposed AutoML Approach}\label{proposed-approach}

\subsection{Problem Formulation}
Given an aggregated energy signal of one house, say $S$, with $n$ samples at $n$ time instances, i.e., the sequence (time series) $[S_1, S_2, ..., S_n]$, the aim of NIALM is to find the $m$ signals of the $m$ appliances inside of the house (i.e., the set $\{s1, s2, ..., sm\}$), that constitute the aggregated signal. Based on that, one should ideally infer the type of appliances inside of the house that correspond to the $m$ individual signals as well as their energy consumption at each time instance, i.e., the set of sequences (time series) $\{[s1_1, ..., s1_n], [s2_1, ..., s2_n], ..., [sm_1, ..., sm_n]\}$. In the following, $\hat{si}_j$ denotes the estimated or predicted value of the power consumption of appliance $i$ at time instance $j$. Also, ${si}_j$ denotes the actual value of the power consumption of appliance $i$ at time instance $j$. This is a supervised ML (regression) problem since our data instances are labeled with numeric values of the total electrical energy consumption of a house. The input variables (ML features) are numeric values of the electrical energy consumption levels of various appliances in the household.

\subsection{Automated ML for NIALM}
Instead of having data scientists perform data preparation, feature selection, model building, and hyper-parameter selection and optimization, AutoML requires the computer itself to carry out these tasks automatically. To our knowledge, AutoML has not yet been applied to the domain of NIALM. Given the fact that we already have various ML algorithms and methods that can be used for NIALM in practice (see Section \ref{literature-survey-benchmarking}), but their exploitation in the industry has not yet been successful, we consider AutoML a vital need for the domain of NIALM.

Therefore, we propose our novel approach based on three pillars: (i) Choosing the best ML model and algorithm for NIALM shall be automated. (ii) Choosing the best hyper-parameters for the model shall be automated. (iii) Supporting novice users by providing them with tips so that they can avoid typical mistakes that can lead to false evaluations. For instance, mixing the training, validation, and testing datasets or shuffling sequential (time-series) data, e.g., through Random cross-validation, must be prohibited.

\begin{figure*}
	\begin{tikzpicture}[
	level 1/.style={sibling distance=20mm},
	edge from parent/.style={->,draw},
	>=latex]
	
	\node[root] {Search Space}
	child {node[level 2] (c1) {Decision Trees}}
	child {node[level 2] (c2) {Random Forests}}
	child {node[level 2] (c3) {GRU}}
	child {node[level 2] (c4) {LSTM}}
	child {node[level 2] (c5) {FCNN}}
	child {node[level 2] (c6) {DAE}}
	child {node[level 2] (c7) {FHMM}}
	child {node[level 2] (c8) {CO}}
	;
	
	\begin{scope}[every node/.style={level 3}]
	\node [below of = c1, xshift=15pt] (c11) {criterion};
	\node [below of = c11] (c12) {min sample split};
	
	\node [below of = c2, xshift=15pt] (c21) {n estimators};
	\node [below of = c21] (c22) {criterion};
	\node [below of = c22] (c23) {min sample split};
	
	\node [below of = c3, xshift=15pt] (c31) {optimizer};
	\node [below of = c31] (c32) {learning rate};
	\node [below of = c32] (c33) {loss function};
	
	\node [below of = c4, xshift=15pt] (c41) {optimizer};
	\node [below of = c41] (c42) {learning rate};
	\node [below of = c42] (c43) {loss function};
	
	\node [below of = c5, xshift=15pt] (c51) {optimizer};
	\node [below of = c51] (c52) {learning rate};
	\node [below of = c52] (c53) {loss function};
	\node [below of = c53] (c54) {n layers};
	\node [below of = c54] (c55) {dropout probability};
	
	\node [below of = c6, xshift=15pt] (c61) {optimizer};
	\node [below of = c61] (c62) {learning rate};
	\node [below of = c62] (c63) {loss function};
	\node [below of = c63] (c64) {sequence length};
	\end{scope}
	
	\foreach \value in {1,2}
	\draw[->] (c1.195) |- (c1\value.west);
	
	\foreach \value in {1,2,3}
	\draw[->] (c2.195) |- (c2\value.west);
	
	\foreach \value in {1,2,3}
	\draw[->] (c3.195) |- (c3\value.west);
	
	\foreach \value in {1,2,3}
	\draw[->] (c4.195) |- (c4\value.west);
	
	\foreach \value in {1,...,5}
	\draw[->] (c5.195) |- (c5\value.west);
	
	\foreach \value in {1,...,4}
	\draw[->] (c6.195) |- (c6\value.west);
	
	\end{tikzpicture}
	\caption{Our proposed model for the search space of the AutoML approach}
	\label{fig:SearchSpace}
\end{figure*}

We build a search space by considering several candidate ML models and hyperparameters for NIALM. Figure \ref{fig:SearchSpace} depicts our proposed model for the search space of the AutoML approach. We consider the following possible values for the hyper-parameters of each ML model in our search space: (i) \textbf{Criterion (see \cite{Sklearn-DT})} $\in$ \{MSE, Friedman\_MSE, MAE\}, (ii) \textbf{Min sample split (see \cite{Sklearn-DT})} $\in$ Uniform [2,200], (iii) \textbf{No. estimators (see \cite{Sklearn-RF})} $\in$ Uniform [5,100], (iv) \textbf{Optimizer (see \cite{Keras-Optimizers})} $\in$ \{Adam, Nadam, RMSprop\}, (v) \textbf{Learning rate} $\in$ \{1e-2,1e-3,1e-4,1e-5\}, (vi) \textbf{Loss function} $\in$ \{mean squared error, mean absolute error\}, (vii) \textbf{No. layers} $\in$ Uniform [5,8], (viii) \textbf{Dropout probability} $\in$ Uniform [0.1,0.6], (ix) \textbf{Sequence length} $\in$ \{64,128,256,512,1024\}

Then, we use Bayesian Optimization to select the best ML model and optimize its hyper-parameters. To this aim, we employ the free open-source Python library \textit{Hyperopt} \cite{hyperopt} for distributed asynchronous hyperparameter optimization. Hyperopt is one of the few libraries that allow search spaces with different types of variables (continuous, ordinal, categorical), varied search scales (e.g., uniform vs. log scaling), and conditional structure (the parameters of one classifier are irrelevant to the other classifiers). Last but not least, similar to our algorithms, this library is also implemented in Python. 

Hyperopt has been designed to accommodate Bayesian optimization algorithms. We deploy the Tree-structured Parzen Estimator (TPE) \cite{Bergstra+2011} algorithm implemented in this library for choosing the best ML model/algorithm and the most adequate hyper-parameters from the search space. In fact, the problem is optimizing a loss function over a graph-structured search space (i.e., configuration space). In our case, we assume a tree structure for the search space. Tree-structured Parzen Estimator (TPE) offers a modeling strategy and scheme for optimizing the criterion of Expected Improvement (EI) for Sequential Model-Based Global Optimization (SMBO) \cite{Bergstra+2011}.

In the search space, we consider the ML approaches shown in Table \ref{tab:sota}, except the Dictionary-Based Approach \cite{ElhamifarSystry2015} as well as the Combined Deep Neural Networks (DNN) and Hidden Markov Models (HMM) approach \cite{MauchYang2016}. We decided to exclude them based on our experiments that showed the gain in terms of the evaluation metrics was not justifiable considering the amount of time and resources they required. Hence, for our automated approach, they do not add enough value. Further, we propose new methods that we find quite suitable for NIALM applications (see Table \ref{tab:extended-benchmarking} in Section \ref{experimental-results}). Those methods are based on Decision Trees, Fully-Connected Neural Networks (FCNN), also known as Multi-Layer Perceptron (MLP), and Gated Recurrent Units (GRU). In Section \ref{experimental-results}, we illustrate our experimental results for the mentioned approaches, thus providing an extended benchmarking. 

\section{Experimental Results}\label{experimental-results}

We name our tool AutoML4NIALM. This is available as free open-source software with a permissive license.

\subsection{Evaluation Metrics}\label{evaluation-metrics}

In this paper, we provide a fair benchmarking using the same datasets, the same evaluation metrics, and the same settings for all existing approaches as well as our proposed methods. The results of this benchmarking are shown in Tables \ref{tab:sota} (only the existing approaches) and \ref{tab:extended-benchmarking} (the extended version). We introduce the deployed evaluation metrics below.

\paragraph{Mean Absolute Error (MAE)}
In our opinion, this is the most relevant metric for typical NIALM applications. The Mean Absolute Error (MAE) for one specific appliance $i$ over $N$ time instances is defined as follows:

\begin{equation}
\text{\textbf{Mean Absolute Error}}_i = \frac{1}{N} \sum_{n=1}^N |si_n-\hat{si}_n|
\end{equation}

Note that we use the average of MAE for all appliances in our benchmarking. Thus, we used the following metric for the results shown on Tables \ref{tab:sota} and  \ref{tab:extended-benchmarking} assuming we have $M$ appliances and $N$ time instances:

\begin{equation}
\text{\textbf{Mean Absolute Error}}_{all} = \sum_{i=1}^M  \frac{1}{N} \sum_{n=1}^N |si_n-\hat{si}_n|
\end{equation}

\paragraph{Disaggregation Accuracy}

Another related evaluation metric is the Disaggregation Accuracy. Note that the definition below clearly shows that this is very different from classification (ML) accuracy despite the similarity in terminology. Again, assuming we have $N$ time instances, the Disaggregation Accuracy for appliance $i$ is defined as follows:

\begin{equation}
\text{\textbf{Disaggregation Accuracy}}_i = 1-\frac{\sum_{n=1}^{N}\left | si_n-\hat{si}_n \right | }{2\sum_{n=1}^{N} \hat{si}_n}
\end{equation}

\paragraph{Classification Accuracy}

This is indeed about the performance of the ML method in terms of the typical ML metric Accuracy. To distinguish this from the Disaggregation Accuracy defined above, we call this one \textit{Classification Accuracy}. However, In the ML literature, this is often simply called \textit{Accuracy}. This metric shows how \textit{accurate} the classification algorithm can predict the actual load (or state) of the appliance. For some appliances like lights, we might consider only two classes, i.e., ON (non-zero load) and OFF (near zero or zero load), whereas, for some other appliances such as dishwashers or microwave ovens, one may consider more than two classes.

\begin{equation}
\text{\textbf{Classification Accuracy}} = \frac{TP + TN}{P + N}
\end{equation}
where TP, TN, P, and N denote the number of True Positive, True Negative, Positive, and Negative data instances. A high classification accuracy can show a good performance of the classifier, but accuracy alone is generally not sufficient to judge that. Therefore, we usually also take a look at other ML metrics, such as Precision and Recall, or a metric called F-Measure or F1-Score that combines both Precision and Recall. The latter is quite useful as we often have a trade-off between the Precision and Recall. Below, we define the said ML metrics.

\paragraph{Precision, Recall and F1-Score}

Precision and Recall illustrate the \textit{Precision} and the \textit{Sensitivity} of the classification algorithm, respectively.

\begin{equation}
\text{\textbf{Precision}} = \frac{TP}{TP + FP}
\end{equation}
where TP and FP denote the number of True Positive and False Positive data instances.

\begin{equation}
\text{\textbf{Recall}} = \frac{TP}{TP + FN}
\end{equation}
where TP and FN denote the number of True Positive and False Negative data instances.

\begin{equation}
\text{\textbf{F1-Score}} = 2 \times \frac{Precision \times Recall}{Precision + Recall}
\end{equation}

\paragraph{Normalized Absolute Distance}

Finally, the Normalized Absolute Distance (NAD) of appliance $i$ over $N$ time instances is defined as follows:
\begin{equation}
\text{\textbf{Normalized Absolute Distance}}_i = \sqrt{\frac{\sum_{n=1}^{N}\left | si_n-\hat{si}_n \right |}{\sum_{n=1}^{N}\left | si_n \right |}} 
\end{equation}

NAD is a metric that we found in the Signal Processing literature. For instance, it is reported in \cite{MauchYang2016}.

\subsection{Experimental Setup}\label{experimental-setup}

To carry out the experiments, we either used the source code provided by the authors or, if that was not possible, carefully reimplemented their approach based on the information provided in the respective papers. We also acknowledge the reuse of algorithms implemented in the free open-source NILM-Toolkit \cite{Kelly+2014}. Moreover, the dataset used for this benchmarking is the REDD dataset \cite{Kolter+2011}. Note that the existing open reference datasets for NIALM have played a key role in making high-impact research in this field possible over recent years and in facilitating our benchmarking. The most well-known and widely used ones are the REDD dataset from the U.S. \cite{Kolter+2011} as well as the UK-DALE \cite{KellyKnottenbelt2015b} and the REFIT datasets from the U.K. \cite{Murray+2017}. Our tool presented in Section \ref{proposed-approach} currently supports the REDD and the UK-DALE data formats out of the box. Nevertheless, we are aware of other existing open datasets for NIALM, such as \cite{Barker+2012,Batra+2013,Makonin+2016}.

Specifically, we used the data of $37$ days from house 1 of the REDD dataset \cite{Kolter+2011} with a sampling rate (frequency) of one per $20$ seconds, i.e., $0.05$ Hz for conducting our benchmarking in Section \ref{literature-survey-benchmarking}, the extended benchmarking in Section \ref{extended-benchmarking}, as well as the experiments on AutoML in Section \ref{automl-experiments}. We also split the dataset into training, validation, and testing segments. Further, in all experiments, we considered $200$ trials of the AutoML algorithm for the optimization and $2000$ maximum number of epochs for the training of the ML model. Also, we set a patience level of $15$. The latter means that even if the validation loss is not decreasing during the training phase, we could still tolerate that for up to $15$ training epochs; thus, the training would not stop immediately. 

In addition, our methodology for conducting the benchmarking is as follows: For each ML algorithm, we consider the Mean Absolute Error (MAE) for disaggregation of the signals of the fridge, lights, sockets, and washer/dryer and calculate the average of the MAE over all the said (groups of) appliances. The highest rank, e.g., in Table \ref{tab:sota}, means the lowest average for MAE over all the said (groups of) appliances, whereas the lowest rank shows the largest MAE on average for them. In Section \ref{extended-benchmarking} below, we explain why we chose MAE among all metrics for the benchmarking.

\subsection{Extended Benchmarking}\label{extended-benchmarking}

We argue that for typical NIALM use cases, e.g., for load balancing in Smart Grids or saving on energy bills for the end users, Mean Absolute Error (MAE) is the best metric to measure the performance of the NIALM approaches. The reason is that, in the end, we are interested in having the least regression error when predicting the loads of appliances. This means the ultimate goal is to reduce the total deviation of the model from the actual power consumption of the target appliance(s). Therefore, that might even imply sacrificing the classification performance (e.g., the Accuracy, Precision, or Recall in predicting the right states/loads of appliances) or other above-mentioned evaluation metrics to achieve that ultimate goal shall not really be an issue. For this reason, we consider only this metric for the benchmarking conducted here.

Table \ref{tab:extended-benchmarking} shows the Mean Absolute Error (MAE) that we observed in our experiments both with the existing state-of-the-art approaches named in Table \ref{tab:sota} and for the proposed approaches based on Decision Trees, Fully-Connected Neural Networks (Multi-Layer Perceptron) and Gated Recurrent Units (GRU). We can see that neither the state-of-the-art approach proposed by Elhamifar and Systry in 2015 (i.e., the Dictionary-Based Approach \cite{ElhamifarSystry2015}) nor the one proposed by Mauch and Yang in 2016 (i.e., the DNN-HMM approach \cite{MauchYang2016}) perform as good as the Deep Neural Network approaches of Kelly and Knottenbelt proposed in 2015 \cite{KellyKnottenbelt2015a}, based on Long Short-Term Memory (LSTM) and Denoising Autoencoders (DAE). 

In addition, we observed through our experiments that neither of the said approaches (Dictionary-Based and DNN-HMM) were efficient in terms of the time and space required for their training. For example, training the Dictionary-Based approach on Commercial Off-The-Shelf (COTS) hardware was so slow in our initial experiments that we decided to substitute the dictionary creation part of that approach with another solution built based on the UCR Matrix Profile \cite{MatrixProfile} for finding similar signals (known as Motifs in Signal Processing).

\begin{table}[ht]
	\centering
	\caption{Extended benchmarking using both the state-of-the-art approaches and proposed algorithms for NIALM}
	\label{tab:extended-benchmarking}
	\begin{tabular}{|p{1cm}|p{4cm}|p{1cm}|}
		\hline
		\textbf{Rank} & \textbf{Algorithm} & MAE \\
		\hline
		1st & \textbf{Decision Trees (DT)} & $14.86$ \\
		\hline
		2nd & \textbf{Gated Recurrent Units (GRU)} & $19.55$ \\
		\hline
		3rd & Long Short-Term Memory (LSTM) \cite{KellyKnottenbelt2015a} & $25.77$ \\
		\hline
		4th & Denoising Autoencoders (DAE) \cite{KellyKnottenbelt2015a} & $41.44$ \\
		\hline
		5th & \textbf{Fully-Connected Neural Networks (FCNN) a.k.a. Multi-Layer Perceptron (MLP)} & $48.96$ \\
		\hline
		6th & Dictionary-Based Approach \cite{ElhamifarSystry2015} & $61.15$ \\
		\hline
		7th & DNN-HMM \cite{MauchYang2016} & $118.52$ \\
		\hline
		8th & Factorial Hidden Markov Models (FHMM) \cite{Reyes-Gomez+2003} & $139.40$ \\
		\hline
		9th & Combinatorial Optimization (CO) \cite{Hart1992} & $204.83$ \\
		\hline
	\end{tabular}
\end{table}

\begin{figure}[h]
	\centering
	\includegraphics[width=0.5\textwidth]{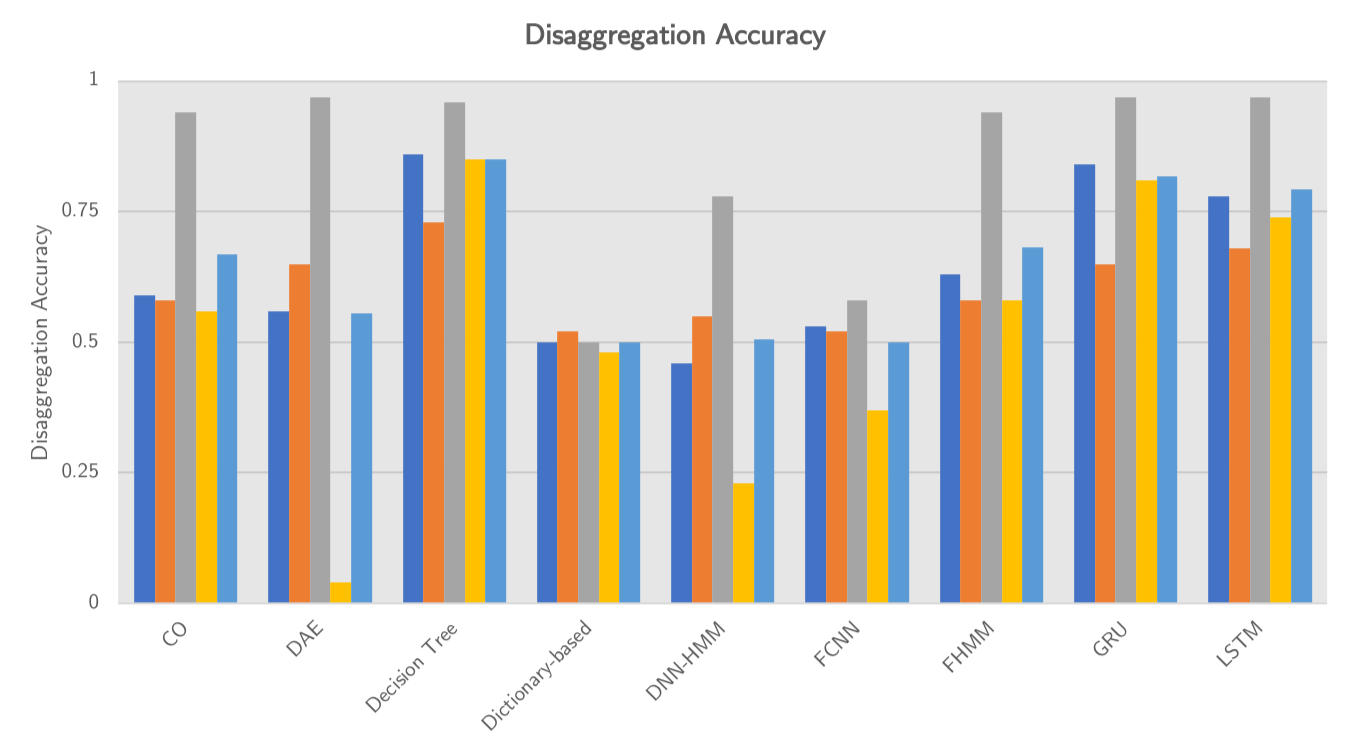}
	\caption{Benchmarking using Disaggregation Accuracy - color codes: dark blue, orange, gray, yellow and light blue represent fridge, lights, sockets, washer/dryer and average for all appliances, respectively.}
	\label{fig:eval-disaggregation-accuracy}
\end{figure}

\begin{figure}[h]
	\centering
	\includegraphics[width=0.5\textwidth]{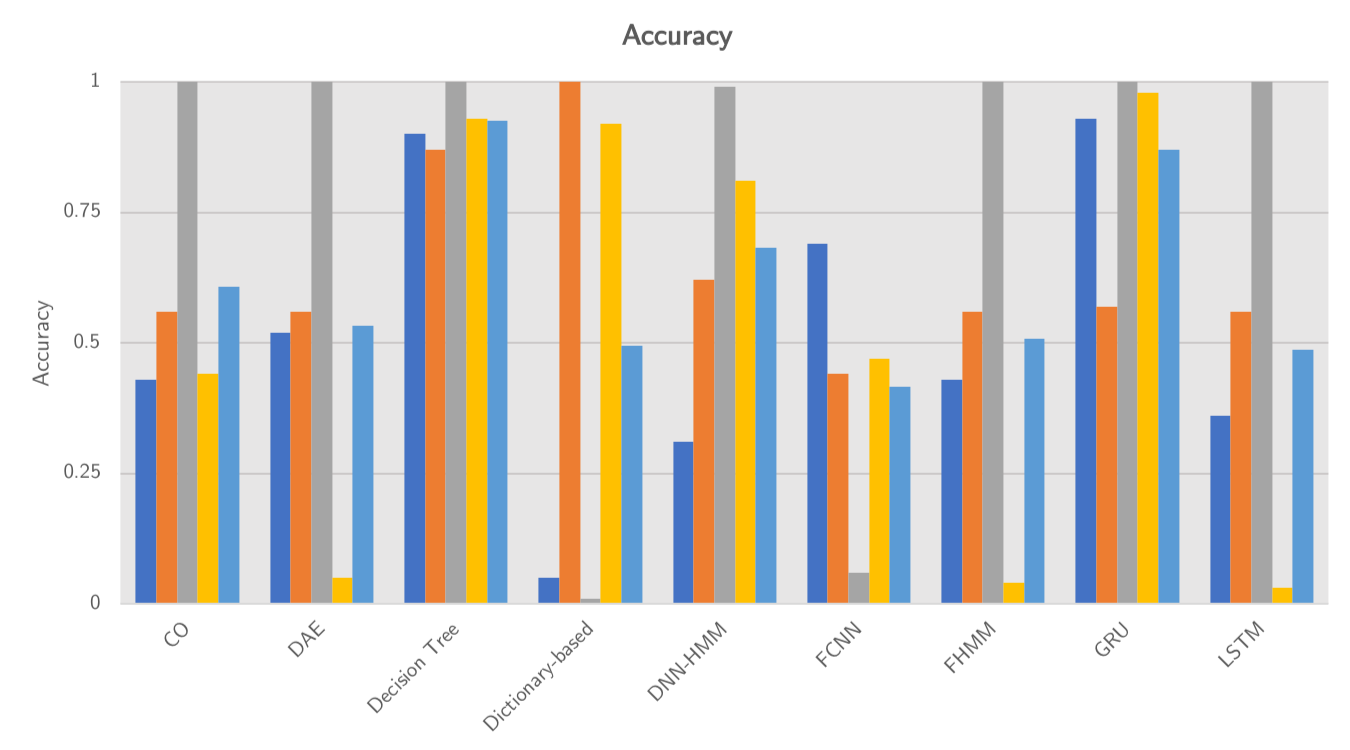}
	\caption{Benchmarking using Classification Accuracy - color codes: dark blue, orange, gray, yellow and light blue represent fridge, lights, sockets, washer/dryer and average for all appliances, respectively.}
	\label{fig:eval-class-accuracy}
\end{figure}

\begin{figure}[h]
	\centering
	\includegraphics[width=0.5\textwidth]{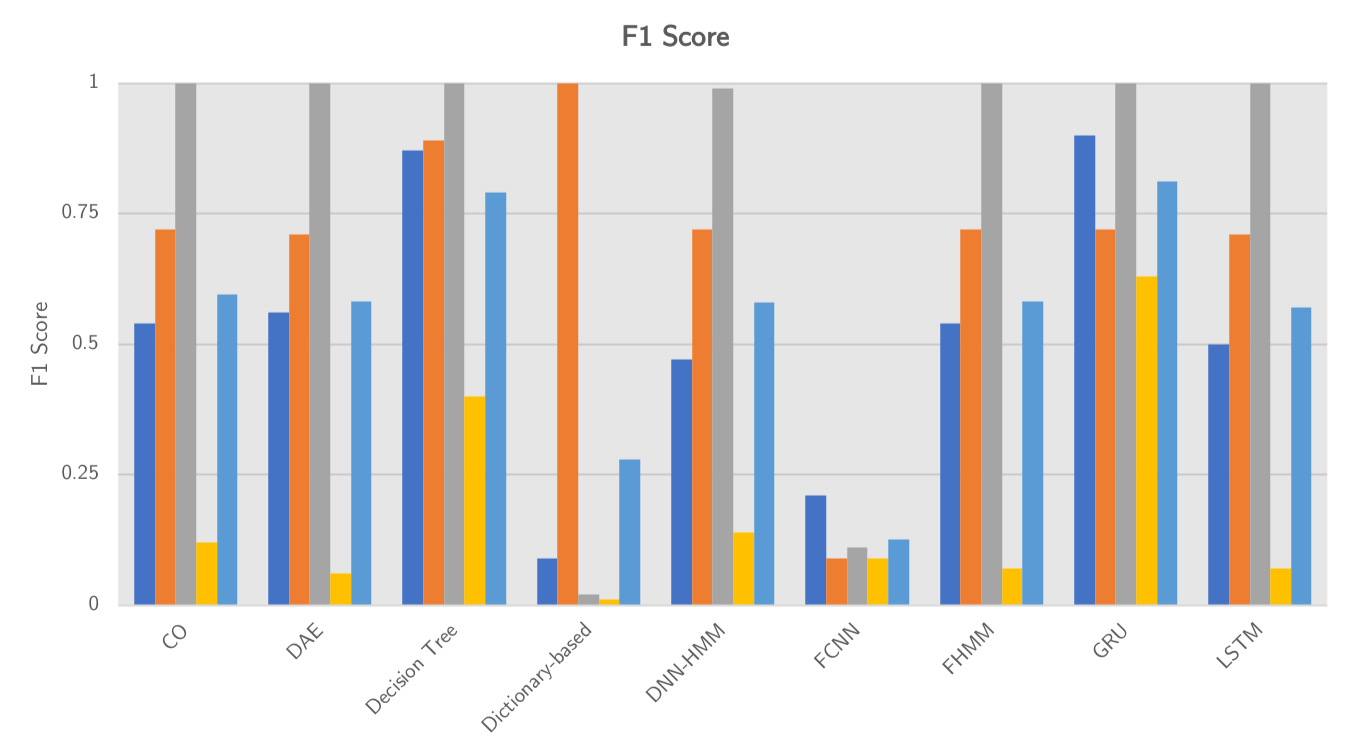}
	\caption{Benchmarking using F1-Score - color codes: dark blue, orange, gray, yellow and light blue represent fridge, lights, sockets, washer/dryer and average for all appliances, respectively.}
	\label{fig:eval-class-F1}
\end{figure}

\begin{figure}[h]
	\centering
	\includegraphics[width=0.5\textwidth]{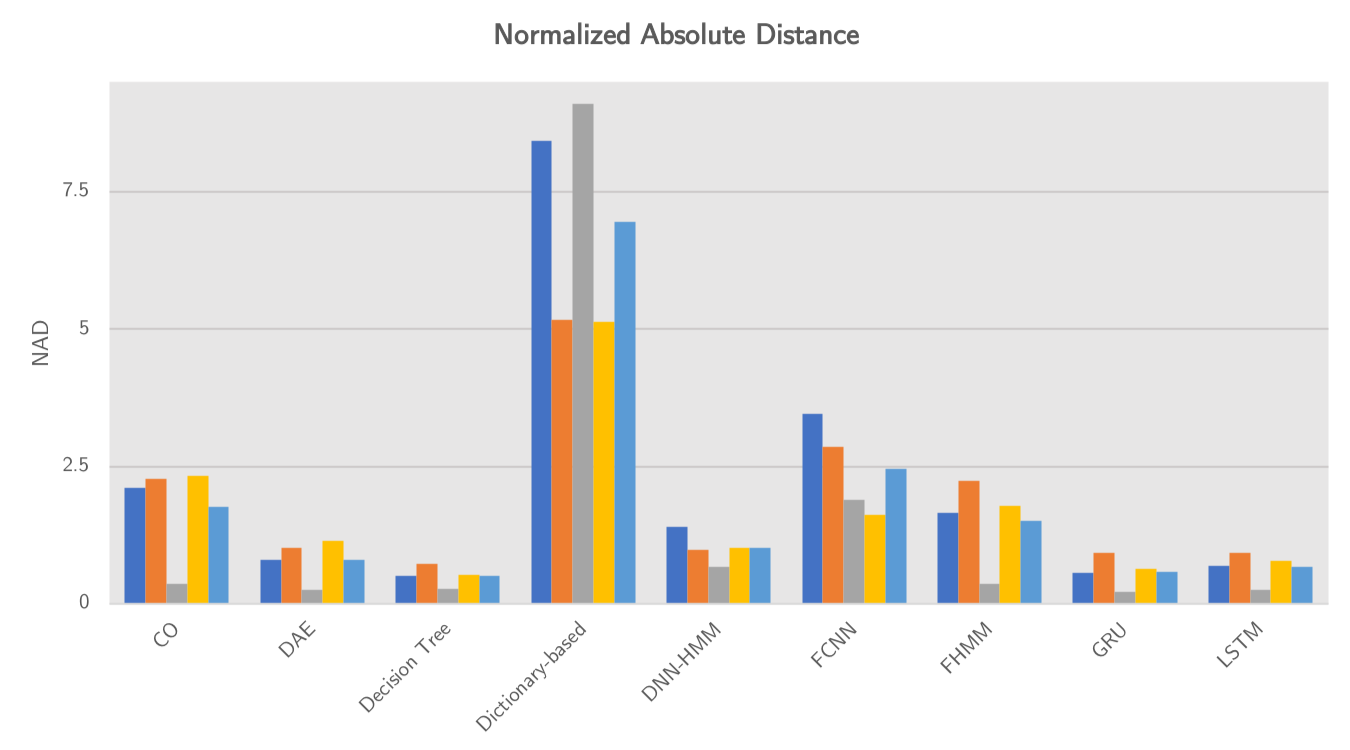}
	\caption{Benchmarking using Normalized Absolute Distance (NAD) - color codes: dark blue, orange, gray, yellow and light blue represent fridge, lights, sockets, washer/dryer and average for all appliances, respectively.}
	\label{fig:eval-class-NAD}
\end{figure}

\begin{figure}[h]
	\centering
	\includegraphics[width=0.5\textwidth]{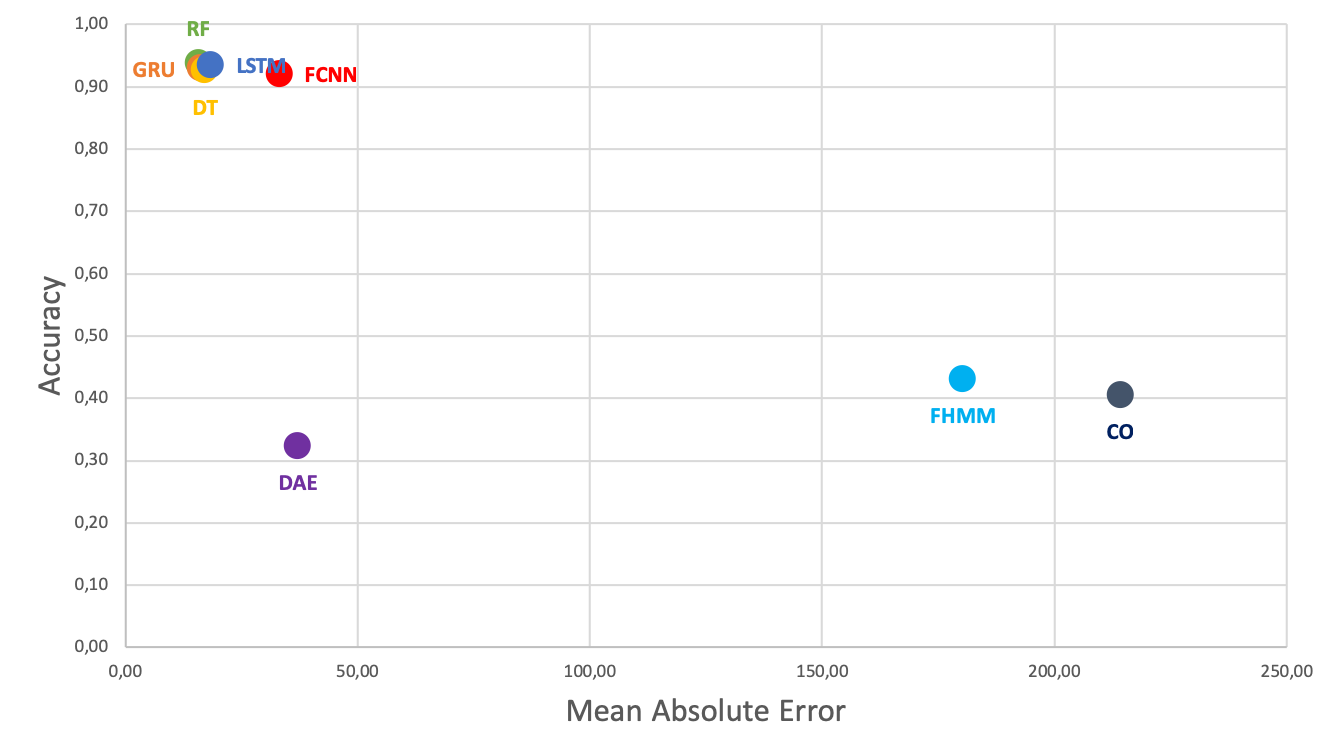}
	\caption{Classification Accuracy vs. Mean Absolute Error (average over all appliances), DT and RF stand for Decision Trees and Random Forests, respectively.}
	\label{fig:ClassAccuracy-vs-MAE}
\end{figure}

Furthermore, Figures \ref{fig:eval-disaggregation-accuracy}, \ref{fig:eval-class-accuracy}, \ref{fig:eval-class-F1}, and \ref{fig:eval-class-NAD} depict our benchmarking results regarding other evaluation metrics, namely Disaggregation Accuracy, Classification Accuracy, F1-Score (i.e., Precision and Recall of classification) and  Normalized Absolute Distance (NAD), respectively. Finally, Figure \ref{fig:ClassAccuracy-vs-MAE} shows the Classification Accuracy as well as the Mean Absolute Error (MAE) for various approaches simultaneously. According to this diagram, Combinatorial Optimization (CO) \cite{Hart1992} and after that, Factorial Hidden Markov Models (FHMM) \cite{Reyes-Gomez+2003} are the worst ones with respect to these two metrics. Also, despite having a low MAE, Denoising Autoencoders (DAE) \cite{KellyKnottenbelt2015a} has the drawback of low classification Accuracy. Finally, we see that there is not much difference between the rest of the approaches when it comes to the mentioned metrics. Thus, choosing the right approach among those remaining ones shall be more a matter of other evaluation metrics or other constraints such as the training time (e.g., obviously very quick in the case of Decision Trees and very long in the case of Deep Neural Networks). Note that on this diagram, we also see an additional algorithm, Random Forest. Actually, we deployed Ensemble Learning to train a number of Decision Trees. However, as illustrated, neither in terms of the MAE nor regarding the Classification Accuracy could we observe any considerable improvement.

\subsection{AutoML Experiments}\label{automl-experiments}

Our proposed Automated Machine Learning (AutoML) approach to NIALM chooses the best ML model, algorithm, and configurations out of the search space that is depicted in Figure \ref{fig:SearchSpace}. Tables \ref{tab:experiments-automl-mae}, \ref{tab:experiments-automl-acc} and \ref{tab:experiments-automl-f1} illustrate the improvements by comparing a classic (i.e., manual) setting with our proposed automated approach based on the Mean Absolute Error (MAE), Classification Accuracy and F1-Score evaluation metrics, respectively.  Interestingly, we see that the AutoML approach even leads to better results by the basic architecture of Neural Networks that we proposed, i.e., Fully-Connected Neural Networks (Multi-Layer Perceptron) compared to Denoising Autoencoders (DAE), not only in terms of the MAE but also in terms of the Classification Accuracy and the F1-Score.

As we saw in Figure \ref{fig:SearchSpace}, we do not have any hyper-parameters for the Factorial Hidden Markov Models (FHMM) \cite{Reyes-Gomez+2003} and Combinatorial Optimization (CO) \cite{Hart1992} approaches. This explains why, in principle, we do not achieve any better results using the AutoML approach for them. Actually, the slightly better MAE shown in Table \ref{tab:experiments-automl-mae} is rather accidental, as we get slightly different results in different runs.

\begin{table}[ht]
	\centering
	\caption{Experimental results for the proposed AutoML approach using MAE}
	\label{tab:experiments-automl-mae}
	\begin{tabular}{|p{1.5cm}|p{1.5cm}|p{1.5cm}|}
		\hline
		\textbf{Algorithm} & Manual & AutoML \\
		\hline
		DT & $21.16$ & $19.05$ \\
		\hline
		GRU & $23.14$ & $19.87$ \\
		\hline
		LSTM \cite{KellyKnottenbelt2015a} & $28.13$ & $21.80$ \\
		\hline
		DAE \cite{KellyKnottenbelt2015a} & \textbf{39.17} & \textbf{35.54} \\
		\hline
		FCNN & \textbf{51.64} & \textbf{34.50} \\
		\hline
		FHMM \cite{Reyes-Gomez+2003} & $118.57$ & $118.57$ \\
		\hline
		CO \cite{Hart1992} & $208.36$ & $204.57$ \\
		\hline
	\end{tabular}
\end{table}

\begin{table}[ht]
	\centering
	\caption{Experimental results for the proposed AutoML approach using Classification Accuracy}
	\label{tab:experiments-automl-acc}
	\begin{tabular}{|p{1.5cm}|p{1.5cm}|p{1.5cm}|}
		\hline
		\textbf{Algorithm} & Manual & AutoML \\
		\hline
		DT & $0.77$ & $0.80$ \\
		\hline
		GRU & $0.84$ & $0.84$ \\
		\hline
		LSTM \cite{KellyKnottenbelt2015a} & $0.51$ & $0.91$ \\
		\hline
		DAE \cite{KellyKnottenbelt2015a} & \textbf{0.65} & \textbf{0.71} \\
		\hline
		FCNN & \textbf{0.37} & \textbf{0.88} \\
		\hline
		FHMM \cite{Reyes-Gomez+2003} & $0.51$ & $0.51$ \\
		\hline
		CO \cite{Hart1992} & $0.61$ & $0.61$ \\
		\hline
	\end{tabular}
\end{table}

\begin{table}[ht]
	\centering
	\caption{Experimental results for the proposed AutoML approach using F1-Score}
	\label{tab:experiments-automl-f1}
	\begin{tabular}{|p{1.5cm}|p{1.5cm}|p{1.5cm}|}
		\hline
		\textbf{Algorithm} & Manual & AutoML \\
		\hline
		DT & $0.61$ & $0.72$ \\
		\hline
		GRU & $0.72$ & $0.74$ \\
		\hline
		LSTM \cite{KellyKnottenbelt2015a} & $0.58$ & $0.82$ \\
		\hline
		DAE \cite{KellyKnottenbelt2015a} &  \textbf{0.59} &  \textbf{0.62} \\
		\hline
		FCNN &  \textbf{0.23} &  \textbf{0.77} \\
		\hline
		FHMM \cite{Reyes-Gomez+2003} & $0.58$ & $0.58$ \\
		\hline
		CO \cite{Hart1992} & $0.59$ & $0.59$ \\
		\hline
	\end{tabular}
\end{table}

\section{Conclusion and Future Work}\label{conclusion-future-work}

In this paper, we proposed a novel approach to enable Automated Machine Learning for the domain of NIALM through Bayesian Optimization. Using our free open-source tool, NIALM domain experts and practitioners could benefit from Machine Learning algorithms and methods without necessarily having data analytics skills. Our tool can automatically choose the best Machine Learning (ML) model and algorithm, the most suitable values for the hyper-parameters of the ML model, and the most adequate configurations for training the ML model. We demonstrated that the proposed AutoML approach indeed improves the quality of the data analytics practices for NIALM regarding various evaluation metrics, such as the Mean Absolute Error (MAE), the Classification Accuracy, and the F1-Score.

Moreover, we conducted a comprehensive survey and a fair benchmarking of the state of the art in NIALM and illustrated that, in many cases, simple and basic Machine Learning models and algorithms, such as Decision Trees, outperform the state of the art. Further, we proposed other novel methods for NIALM. In addition, a vital lesson learned from the conducted study is that there is no one-size-fits-all approach to NIALM. In other words, various state-of-the-art machine learning approaches could offer great results depending on the appliance of interest and the evaluation metric of choice. Hence, there is no point in pushing the user towards a certain approach and prescribing that as the best one for all appliances and all purposes. Instead, we believe that the user shall be able to benefit from the automation of the data analytics tasks via tool support to benefit from all state-of-the-art approaches.

Finally, as future work, we plan to extend our open-source tool with more algorithms and methods as well as more features for supporting and guiding users through our Graphical User Interface (GUI). Also, we may apply our AutoML approach to other domains beyond NIALM. Further, one may think of other optimization algorithms for AutoML other than the one we used here for Bayesian Optimization. In addition, currently, the input data has to be in the format used by the two major existing open reference datasets for NIALM. However, one could possibly increase user convenience by supporting other formats as well.

\section*{Acknowledgment}
This work is partially funded by the German Federal Ministry of Education and Research (BMBF) through the Software Campus 2.0 project ML-Quadrat.
This preprint has not undergone any post-submission improvements or corrections. The Version of Record of this contribution will be published in a book chapter by Springer and will become available online.

\bibliographystyle{IEEEtran}
\bibliography{refs}

\end{document}